\documentclass{article}

\begin{document}

\title{How can an electron interact with itself?}

\author{Yong Gwan Yi}

\maketitle

\bigskip
\bigskip

The self-energy of an electron is an old problem 
of classical electrodynamics. It assumes an 
interaction of an electron with the field that the 
electron produces. The quantum theory of electron 
has been developed with the discovery of positron, 
putting the problem of self-energy in a critical 
state. I would like to discuss its causal effect 
that can be assumed in the modern repetition of the 
old problem. 

The quantum theory of electron began with the 
Dirac equation, which allows for electrons having 
equally either positive or negative energy as 
solutions [1]. Dirac assumed in the vacuum the 
negative energy states completely filled with 
electrons. Later, it was assumed that a 
vacated negative energy state appears as a 
positive energy electron of charge $|e|$. 
According to Dirac, an electron-positron pair 
production could be explained as due to the 
transition from a negative to a positive energy 
state of an electron. Feynman, and earlier 
Stueckelberg, proposed an interpretation 
of the positron as a negative energy electron 
moving backward in time [2]. An electron-positron 
pair production and annihilation could thus be 
visualized as a closed loop by a positive and a 
negative energy electrons, characterizing the 
vacuum polarization.

Weisskopf wrote in retrospect, Pauli asked me to 
calculate the self-energy of an electron according 
to the positron theory [3]. Weisskopf calculated 
the difference between the energy of the vacuum 
and the energy of an electron in the vacuum [4]. 
As noted by Furry, the calculation results in a 
logarithmic divergence of self-energy. In the 
positron theory, an electron in the vacuum causes 
a considerable change in the distribution of the 
vacuum electrons, and the broadening of the charge 
distribution makes the electrostatic self-energy 
to diverge logarithmically. 

Feynman put forward an intelligible method of using 
diagrams [5]. It provides an intuitive way of 
arriving at the correct result with the aid of a 
convergence factor
\begin{equation}
\frac{1}{q^2}\Rightarrow\frac{1}{q^2}\biggl(
\frac{-\Lambda^2}{q^2-\Lambda^2}\biggr)
\quad\mbox{or}\quad
\frac{1}{q^2}-\frac{1}{q^2-\Lambda^2}.
\end{equation}
The convergence factor is adopted only to define 
the formally divergent solution. Its physical 
meaning remains unexplained [6]. The Feynman 
diagram itself is also unnatural. In the classical 
theory, the self-force is the reaction back on the 
electron of its own radiation field. This is taken 
over into the Feynman diagram, representing the 
self-energy due to the emission and absorption of 
a virtual transverse photon. The Feynman diagram 
for the self-energy is no more than a quantum 
theoretical description of the static self-force. 
Note in the diagram a complete neglect of the velocity 
of propagation.

The following conclusion can be drawn from the 
qualitative argument of Weisskopf's calculation: 
The self-interaction is an effect on the electron 
of static polarization induced in the vacuum 
electrons due to the presence of an electron. If the 
photon virtually disintegrates into an 
electron-positron pair and annihilates again into a 
photon for a certain fraction of time, the electron 
closed loop gives an additional correction to the 
photon propagator through which an electron interacts 
with itself. The modification of the photon 
propagator is then the replacement
\begin{equation}
\frac{1}{q^2}\Rightarrow\frac{1}{q^2}
[-\Pi_{\mu\nu}(q)]\frac{1}{q^2}.
\end{equation}
The polarization tensor $\Pi_{\mu\nu}(q)$ is 
written as the sum of a constant term $\Pi(0)$ and 
additional terms. The leading term is a positive 
constant that depends quadratically on $\Lambda$. 
Using the operator relation, approximately, we 
write the replacement in the form
\begin{equation}
\frac{1}{q^2}\Rightarrow\frac{1}{q^2}-\frac{1}
{q^2-\Pi(0)}.
\end{equation}
It becomes evident that the vacuum polarization 
effect gives rise to a modification of the photon 
propagator corresponding to the convergence factor. 
By including the electron closed loop in the photon 
propagator, moreover, the Feynman diagram for 
self-energy can be put into accord with the positron 
theory. Thus, the conclusion can also be drawn from 
the diagrammatic argument of Feynman's calculation. 
An electron may interact with itself through the 
vacuum polarization induced in the vacuum electrons 
due to the electron. 

The vacuum polarization is calculated to diverge 
quadratically. Feynman has used a convergence 
procedure found by Bethe, and Pauli and Villars [7]. 
It takes a difference between the closed loops for 
electrons of mass $m$ and of mass 
$(m^2+\lambda^2)^{1/2}$, and integrates over 
$\lambda$ to $\Lambda$. In the form of expression, 
it is just like the convergence factor for photon in 
terms of the electron closed loop. When viewed from 
the present point, therefore, the procedure is 
identified with the replacement
\begin{equation}
-\Pi_{\mu\nu}(q)\Rightarrow\frac{1}{q^2}
[-\Pi_{\mu\nu}(q)]\frac{1}{q^2},
\quad\mbox{so}\quad
\frac{1}{q^2}-\frac{1}{q^2-\Pi_{\mu\nu}(q)}.
\end{equation}
This shows the vacuum polarization that starts and 
ends with photon. In the explicit form, it suggests to 
calculate the vacuum polarization in the production and 
annihilation, or in the cause and effect, rather than 
the vacuum polarization itself. The integration over
$\lambda$ can be considered as that for the negative 
energy states. The convergence procedure may be explained 
from this point of view.

\end{document}